\def\twocol{2}
\def\onecol{1}
\def\colopt{\onecol}
\ifnum\colopt=\twocol
 \documentstyle[floats,twocolumn,aps,psfig]{revtex}
 \def\Section#1{}
\else
 \documentstyle[floats,preprint,aps,psfig]{revtex}
 \def\Section#1{\section{#1}}
\fi

\def\cd{c^\dagger}

\def\tp{{t'}}
\def\abs#1{\vert #1 \vert}
\def\beq{\begin{equation}}
\def\eeq{\end{equation}}
\def\bea{\begin{eqnarray}}
\def\eea{\end{eqnarray}}
\def\nn{\nonumber}
\def\ie{{\it i.e.}}
\def\up{\uparrow}
\def\dn{\downarrow}
\def\nc{n}
\font\amsmath=msbm10 scaled \magstep1

\def\Zed{\hbox{\amsmath Z}}

\begin{document}
\tolerance 50000
\preprint{
\begin{minipage}{5in}
\rightline{\small cond-mat/9908398,
La Plata-Th 99/09, LPENS-Th-17/99,}
\rightline{\small ETH-TH/99-20, SISSA 103/99/EP}
\end{minipage}
}

\draft
\ifnum\colopt=\onecol
 \tightenlines
\fi

\ifnum\colopt=\twocol
 \twocolumn[\hsize\textwidth\columnwidth\hsize\csname @twocolumnfalse\endcsname
\fi

\title{
Doping-dependent magnetization plateaux in $p$-merized Hubbard
chains\footnote{Work done under partial support of the EC TMR Programme
{\em Integrability, non-per\-turba\-tive effects and symmetry in
Quantum Field Theories}.}}
\author{D.C.\ Cabra$^{1}$, A.\ De Martino$^{2}$,
A.\ Honecker$^{3}$\footnote{A Feodor-Lynen fellow of the Alexander von
Humboldt-foundation.}\footnote{Present address: Institut f\"ur Theoretische
Physik, TU Braunschweig, 38106 Braunschweig, Germany.},
P.\ Pujol$^{2}$ and P. Simon$^{4}$}
\address{
$^{1}$Departamento de F\'{\i}sica, Universidad Nacional de la Plata,
      C.C.\ 67, (1900) La Plata, Argentina.\\
Facultad de Ingenier\'\i a, Universidad Nacional de Lomas de
Zamora, Cno. de Cintura y Juan XXIII,
(1832) Lomas de Zamora, Argentina.\\
$^{2}$Laboratoire de Physique\footnote{URA 1325 du CNRS associ\'ee
   \`a l'Ecole Normale Sup\'erieure de Lyon.}
      Groupe de Physique Th\'eorique 
      ENS Lyon, 46 All\'ee d'Italie, 69364 Lyon C\'edex 07, France.\\
$^{3}$Institut f\"ur Theoretische Physik, ETH-H\"onggerberg,
     8093 Z\"urich, Switzerland. \\
$^{4}$International School for Advanced Studies,
     Via Beirut 2-4, 34014 Trieste, Italy.\\
}

\date{February 14, 2000}
\maketitle
\begin{abstract}
\begin{center}
\parbox{15cm}{We study zero-temperature Hubbard chains with periodically
modulated hopping at arbitrary filling $\nc$ and magnetization
$m$. We show that the magnetization curves have plateaux at
certain values of $m$ which depend on the periodicity $p$ and the
filling. At commensurate filling $\nc$ a charge gap opens and then
magnetization plateaux correspond to fully gapped situations.
However, plateaux also arise in the magnetization curves at fixed
$\nc$ between the commensurate values and then the plateau-value
of $m$ depends continuously on $\nc$ and can thus also become
irrational. In particular for the case of dimerized
hopping ($p=2$) and fixed doping we find that a plateau appears at
$m = 1- \nc$. In this case, there is still a gapless mode on the
plateau leading to thermodynamic behavior which is different from
a completely gapped situation.}
\end{center}
\end{abstract}

\pacs{
\hspace{-13mm}
}
\ifnum\colopt=\twocol
 \vskip1pc]
\fi

\section{Introduction and summary of results}

Macroscopic quantum phenomena in strongly correlated electron systems
in low dimensions are presently the subject of intense research. In
particular, plateaux in magnetization curves of quantum magnets
have recently received much attention, a central observation being
that the plateaux occur at (typically simple) {\it rational} fractions
of the saturation magnetization. In one dimension, this is by now rather well
theoretically understood in terms of a quantization condition
that involves the volume of a translationally invariant unit cell
(see e.g.\ \cite{AOY,Totsuka,weSpin,poly}).
Various materials exhibiting plateaux have also been studied 
during the past few years, including e.g.\ a new candidate for a frustrated
trimerized chain material \cite{trimer}. One of the clearest examples
is given by the low-temperature magnetization process of NH$_4$CuCl$_3$
\cite{NH4CuCl3} where one observes plateaux with $1/4$ and $3/4$ of
the saturation magnetization thus demonstrating rationality even if
it is theoretically still unclear why precisely those two numbers are
observed in NH$_4$CuCl$_3$.

The purpose of the present letter is to start a systematic
investigation of the effect of doping on the quantization condition for
the appearance of magnetization plateaux. We will show with examples
that plateaux can appear in doped systems at magnetization values
which depend continuously on the doping and are thus in general
{\it irrational}. This observation itself is not entirely new, but
we believe our physical interpretation in terms of one class of excitations
(say the up electrons) being pinned at commensurate filling while the filling
of a different class of excitations (the down electrons) remains
adjustable to be new. This interpretation also suggests that doping-dependent
magnetization plateaux are probably a generic phenomenon.
One of the known cases is the
one-dimensional Kondo lattice model \cite{KLM} where unpaired
spins behave ferromagnetically thus giving rise to a spontaneous
magnetization of a value controlled by doping. The other example
is an integrable spin-$S$ generalization of the $t-J$ chain doped
with $(S-1/2)$ carriers \cite{FrSo} where, however, the appearance
of plateaux is restricted to large magnetization values.

As a first step towards a general understanding we study the effect
of both a magnetic field and a
periodic modulation of the hopping amplitude ($p$-merization) on a
{\it doped} one-band Hubbard chain. One simple motivation is that
for pure spin systems (corresponding to a half-filled Hubbard
model) in a magnetic field, $p$-merized Heisenberg chains have
turned out to be among the simplest examples \cite{poly}. In
addition, structural modulations can give rise to dimerized
coupling constants e.g.\ in chains and ladders \cite{DR}, the
organic (super)conductors \cite{orgS} (some of which come
naturally at quarter filling) and the ferroelectric perovskites
\cite{EIT}.

The one-band Hubbard model is not only used to describe realistic
situations, but at least in one dimension it is also a useful model
for technical reasons.
For example, the Hubbard chain is exactly solvable by means of the
Bethe Ansatz (BA) \cite{LW}. This can be used as an input to a
bosonization analysis \cite{FK,EF,PS} in order to study
generalizations of this model (see e.g.\ \cite{gen2}).

In this letter we show that there are several different phases
in the $\mu$-$h$ plane 
(see Fig.~\ref{schem}
for a schematic illustration of the cases $p=2$ and $p=3$; abbreviations
refer to the corresponding regions of this figure):

\begin{figure}[bt]
\centerline{\psfig{figure=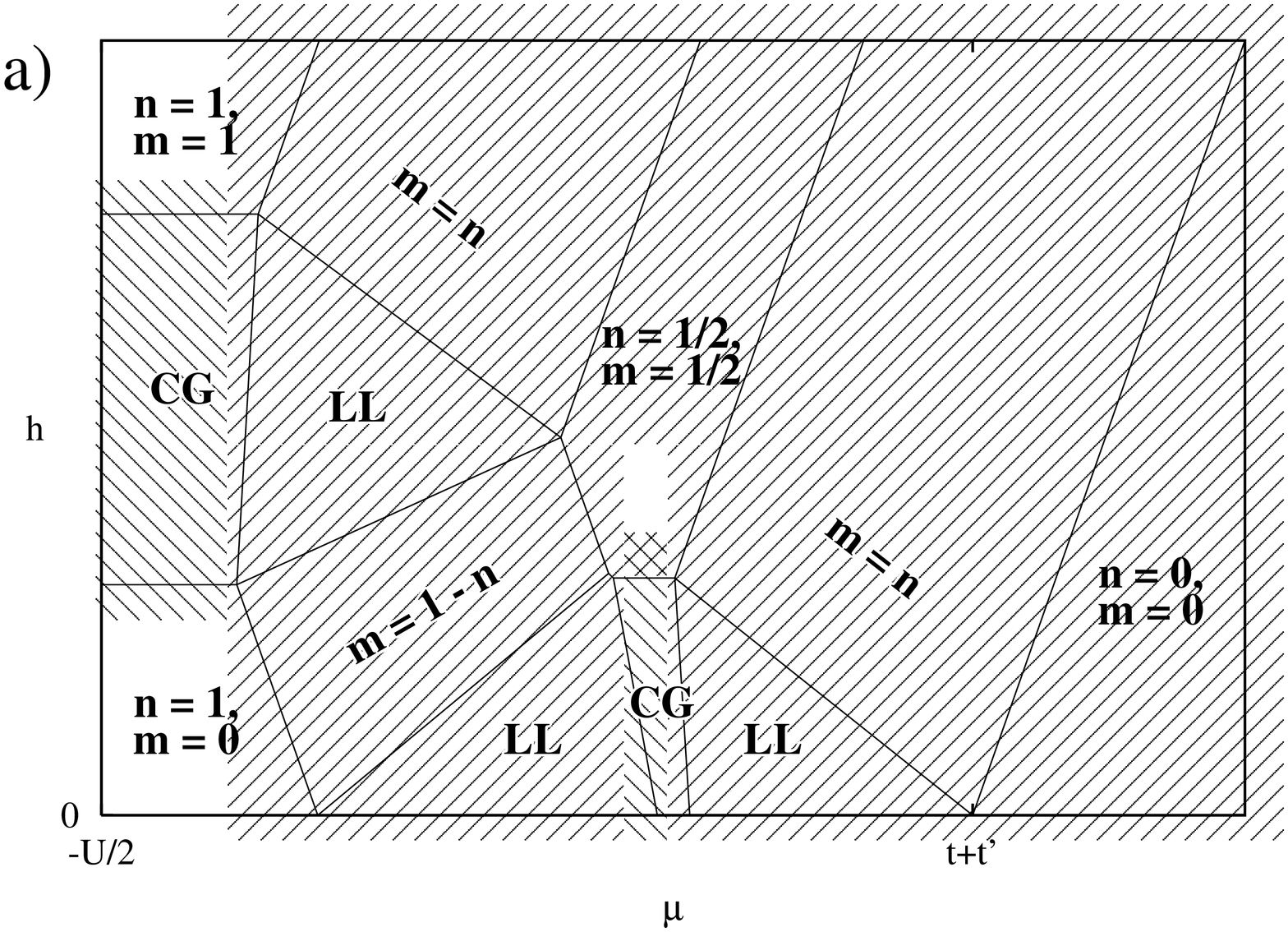,width=\columnwidth,angle=270}}
\centerline{\psfig{figure=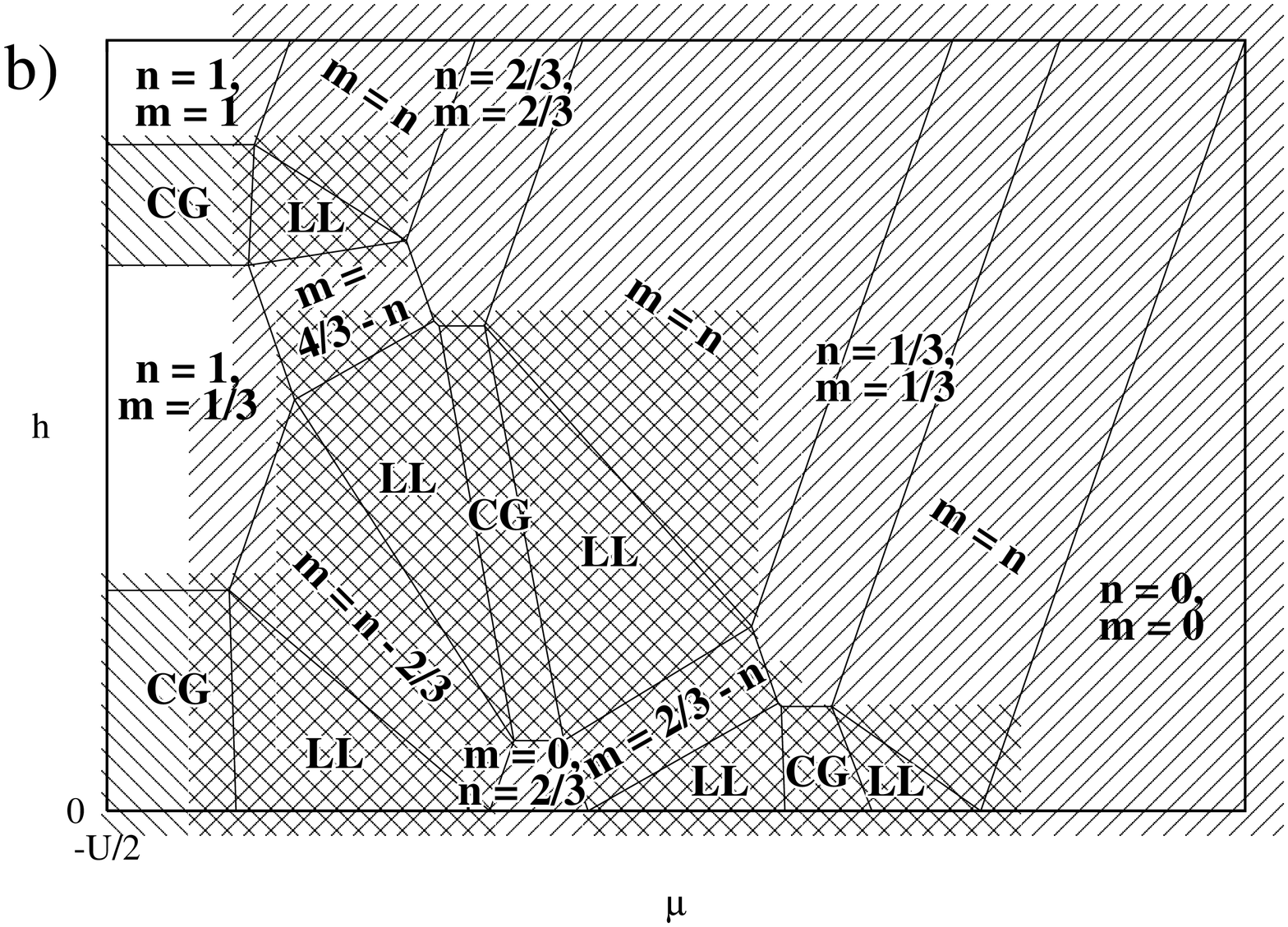,width=\columnwidth,angle=270}}
\smallskip
\caption{
Schematic groundstate phase diagram of a) the dimerized Hubbard
chain ($p=2$) and b) the trimerized chain ($p=3$).
For explanations compare the text.
\label{schem}
}
\end{figure}

\begin{itemize}
\item[i)] If both quantization conditions\footnote{$\nc$ and $m$ are
normalized such that $0 \le \nc \le 2$ and $\abs{m} \le 1$.}
\beq
{p \over 2} \left(\nc \pm m \right) \in \Zed
\label{cond}
\eeq
are satisfied, both spin and charge excitations are gapful and
hence all correlators decay exponentially at large distances
(regions labeled by fixed $n$ and $m$ in Fig.~\ref{schem}).
In this case, we find plateaux in the magnetization curves since
the presence of a spin gap is equivalent to the appearance of a plateau.
\item[ii)] A charge gap (`CG') can open if the combination $p \nc \in \Zed$
of the conditions (\ref{cond}) is satisfied. This includes the well-known
charge gap at half filling ($\nc = 1$) and also the charge gap in
the quarter-filled ($\nc = 1/2$) dimerized Hubbard chain ($p=2$) \cite{MP}.
\item[iii)] If only one of the conditions (\ref{cond}) is fulfilled
(solutions are indicated in the corresponding regions of Fig.~\ref{schem})
and in addition doping $\nc$ is kept fixed, a magnetization plateau opens,
{\it but one mode remains gapless} (similar observations have been made in
other systems \cite{weSpin,FrSo,Tan,Kagome}). In contrast to the
gapful magnetic behavior, charge transport remains metallic in this
phase. \\
The existence of these plateau phases is the main result of the
present letter. A particularly appealing
aspect of the plateaux predicted here is that they can appear at low
magnetization (and thus at small magnetic fields) if the doping is
chosen suitably.
\end{itemize}
In the remaining cases, both spin and charge sectors are massless,
leading to a Luttinger liquid (`LL').

\section{Hamiltonian and strong-coupling limit}

To be concrete, we study the following model Hamiltonian
\bea
H &=& - \sum_{x=1}^L t(x) \sum_{\sigma}
   \left(c^{\dagger}_{x+1,\sigma} c_{x,\sigma}
    + \cd_{x,\sigma} c_{x+1,\sigma}\right) \nn \\
 && + U \sum_{x=1}^L n_{x, \up} n_{x,\dn}
    + \mu \sum_{x=1}^L \left(n_{x, \up} + n_{x,\dn}
    \right) \nn \\
 && -{h \over 2} \sum_{x=1}^L \left(n_{x, \up} -
 n_{x,\dn}\right)
\label{pHub}
\eea
where $t(x) = t$ if $x \neq mp$ and $t(mp) = \tp = t + \delta$. Here,
$c^{\dagger}$ and $c$ are electron creation and annihilation
operators, $n_{x,\sigma} = c^{\dagger}_{x,\sigma} c_{x,\sigma}$
the number operator
and $\sigma = \uparrow,\downarrow$. $\mu$ is the chemical
potential and $h$ is a magnetic field.

First we consider the limit $\tp = 0$. Then
the chain (\ref{pHub}) decouples into clusters of $p$ sites and one
can use simple arguments in the spirit of \cite{weSpin,poly}:
The number of up and down electrons on a $p$-site cluster
must both be integer which is equivalent to imposing both
conditions (\ref{cond}).
All these states are clearly fully gapped at $\tp = 0$. Thus, they
will remain fully gapped if one switches on a small
perturbation $\tp > 0$, only the transitions between these fully
gapped states will soften. In fact, we will argue soon that these
fully gapped states survive even until $\tp = t$.

\section{Bosonization}

Now we turn to a bosonization analysis starting with $h=0$ (for a
related recent study of a dimerized Hubbard chain see \cite{FGN}).
For $\tp = t$ ($\delta = 0$) and $\nc \ne 1$, the Hubbard chain
(\ref{pHub}) can be represented by two bosonic fields with
Hamiltonian \cite{FK}
\beq \sum_{i= c,s} {v_i \over 2} \int dx ~
\left[ \left( \partial_x \phi_i \right)^2 + \left( \partial_x
\theta_i \right)^2 \right] ~,
\label{contham}
\eeq
where $\phi =
\phi_R + \phi_L$ and $\theta = \phi_R - \phi_L$. The spin and
charge fields are given by $\phi_s = {1 \over \sqrt{2}} (
\phi_{\uparrow} - \phi_{\downarrow})$ and $\phi_c = {1 \over \xi }
( \phi_{\uparrow} + \phi_{\downarrow})$, where
$\phi_{\uparrow,\downarrow}$ are
compactified with periodicity $\phi_{\uparrow,\downarrow}
\rightarrow \phi_{\uparrow,\downarrow} + \sqrt{\pi} ~ \Zed $. The
parameter $\xi$ and the Fermi velocities $v_i$ can be obtained
exactly from the BA solution of the model in terms of $U$ and
$\mu$.

If we now turn on a non-zero but small $\delta$, one can show using the
continuum representation of creation and annihilation operators in
terms of the bosonic fields that the most relevant perturbations
to the free Hamiltonian are given by
\beq
O_{pert} =
 \lambda \sin [k_+ /2 + p k_+ x - \sqrt{\pi} \xi \phi_c ]
   \times \cos [\sqrt{2\pi} \phi_s ]
+ \lambda' \cos( k_+ + 2 p k_+ x -  \sqrt{4 \pi} \xi \phi_c  ) \, ,
\label{pert}
\eeq
where $\lambda$ and $\lambda'$ are proportional to $\delta$ and
$k_+ = k_{F,\uparrow} + k_{F,\downarrow} = \pi \nc$.
We can now study the values of $\nc$ for which these operators
are commensurate. In this way, if $p \nc \in \Zed$ the $\lambda'$
term appears as a relevant perturbation, opening a charge gap.
If we restrict further to $p \nc /2 \in \Zed$
the $\lambda$ term is present as well and opens also a spin gap.

We consider now the more complicated case of non-zero
magnetization. The Hubbard chain in the presence of a magnetic
field is also integrable \cite{FK,EF}. The large-scale behavior
of the system is given by a two-field Hamiltonian like
(\ref{contham}), but now the effective charge and spin fields are
given by \cite{PS}

\beq
\left(\matrix{ \phi_c \cr \phi_s} \right) = {1 \over \det~Z }~
\pmatrix{ Z_{ss} & Z_{ss} - Z_{cs} \cr
Z_{sc} & Z_{sc} - Z_{cc}}
\left( \matrix{ \phi_\uparrow \cr \phi_\downarrow } \right) ~,
\label{change1}
\eeq
where $Z$ is the so-called charge matrix
given in \cite{FK}, whose entries can be obtained exactly from
the BA solution in terms of $U$, $\mu$ and $h$. These quantities
determine the scaling dimensions of the vertex operators.

For small $\delta$ the perturbing operators read
\bea
O_{pert} &=&
 \lambda \sin [k_+ /2 + p k_+ x - \sqrt{\pi} \left( Z_{cc}\phi_c -
Z_{cs} \phi_s \right) ] \nn\\
&& \times \cos [k_- /2 + p k_- x \nn \\
&& \quad - \sqrt{\pi} \left( (Z_{cc}-2Z_{sc})\phi_c -
(Z_{cs}-2Z_{ss})\phi_s \right)]
\nn\\
&+& \lambda' \cos( k_+ + 2 p k_+ x
-  \sqrt{4\pi} ( Z_{cc} \phi_c - Z_{cs} \phi_s ) ) ~,
\label{pert2}
\eea
where now $k_- = k_{F,\uparrow} - k_{F,\downarrow} = \pi m$.
If all the operators are commensurate both degrees of freedom
are massive since the perturbing operators are relevant.
We thus have a magnetization plateau with a charge gap.
This is achieved when the two conditions (\ref{cond})
are simultaneously satisfied, which generalizes the result
for $p$-merized spin chains \cite{poly} to $\nc \ne 1$.

When only one of these conditions is
satisfied, say $p \left(\nc + m \right)/2 \in \Zed$,
the Hamiltonian can be written as
\bea
H &=& \int dx ~ {v_\uparrow \over 2} ~
\left[ \left( \partial_x \phi_\uparrow \right)^2 + \left(
\partial_x \theta_\uparrow \right)^2 \right] \nn \\
&&+ {v_\downarrow \over 2} ~ \left[ \left( \partial_x \phi_\downarrow
\right)^2 + \left( \partial_x \theta_\downarrow \right)^2 \right]
+ \lambda \sin 2 \sqrt{\pi} \phi_\uparrow ~.
\label{hamk}
\eea
Terms mixing derivatives of the up and down
fields, which come from the $U$ interaction,
can be shown to be irrelevant using a treatment similar to Ref.\ \cite{gen2}.
One can integrate out the massive field $\phi_\uparrow$ and obtain
an effective large scale Hamiltonian for $\phi_\downarrow$,
with an effective Fermi velocity and Luttinger parameter $K$.
This field is apparently massless but constrained to be in a
particular topological sector if one imposes fixed filling (which seems to be
natural from the experimental point of view).
In a system of finite size $L$, this constraint takes the form
$L \nc = \frac{1}{\sqrt{\pi}} ( \phi_\uparrow + \phi_\downarrow ) |_0^L$
and allowing only small energy fluctuations locks the quantity
$Q= {1 \over \sqrt{\pi}} \phi_\downarrow |_0^L$ to be also constant
(see \cite{prep} for more details).

If this constraint were not imposed, $Q$ would vary and the
susceptibility would then have the standard form
$\chi ={K / (2 \pi v)}$. On the other hand, if we impose
$Q_0 = L ( \nc - m )/2 $, we obtain
$\chi = {1 \over \beta L} \left( \langle Q^2\rangle
- \langle Q\rangle^2 \right) = 0 $, indicating the presence
of a magnetization plateau.
Since the  local part of the down sector in (\ref{hamk}) remains
massless, local correlation functions of the spin-down field
decay algebraically and one can also show that the low-temperature
specific heat is linear in the temperature.
This situation is interesting since it combines a massless behavior
with a magnetization plateau at $p \left(\nc + m \right)/2 \in \Zed$
for arbitrary values of $\nc$.

\section{Small on-site repulsion}

A complementary derivation of these doping-de\-pendent magnetization
plateaux (at fixed $\nc$) is given by the following discussion
of the limit of small interaction $U$.

\begin{figure}[bt]
\centerline{\psfig{figure=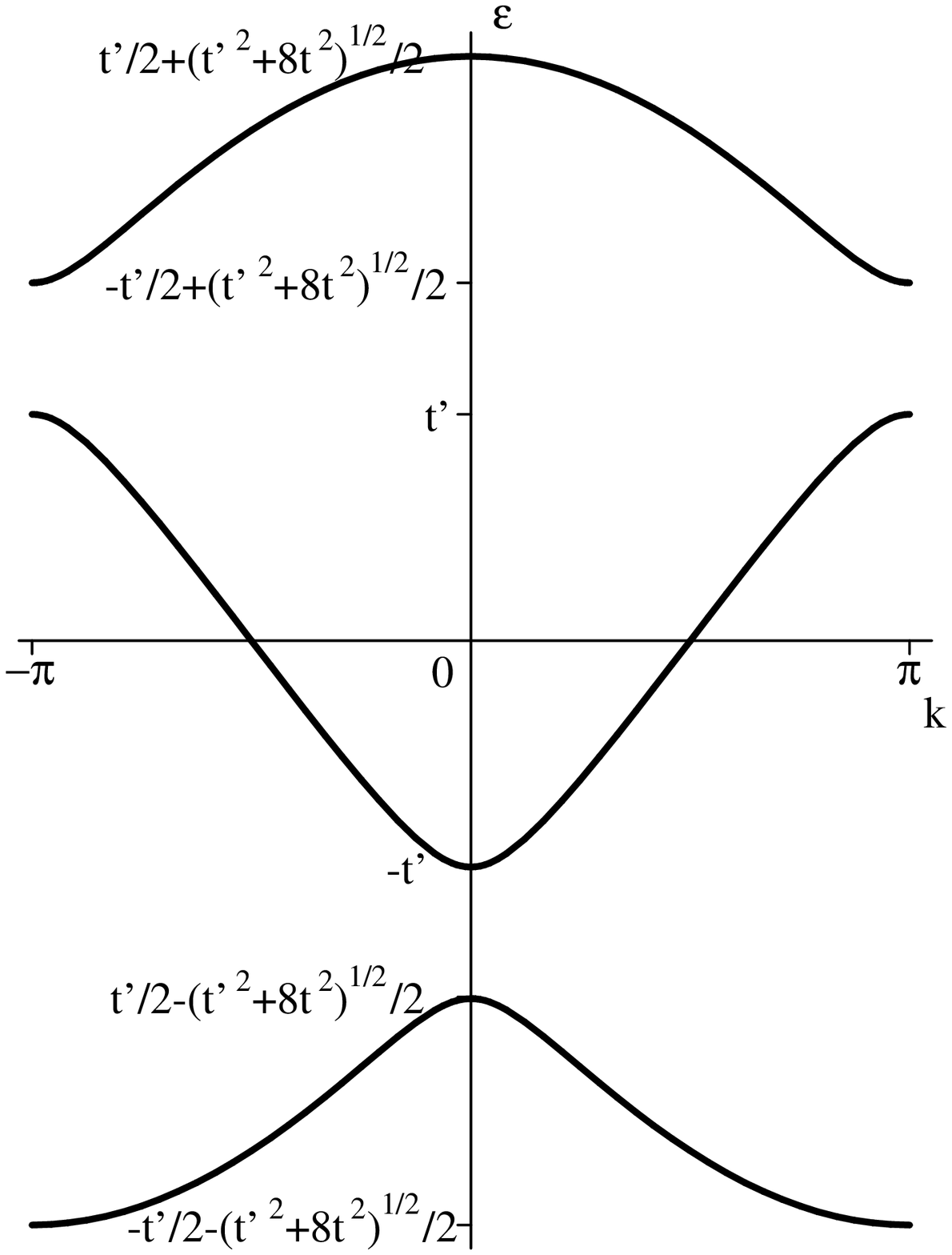,width=0.6\columnwidth,angle=0}}
\smallskip
\caption{
Band structure of the trimerized chain ($p=3$) at $U=0$ for
$\tp < t$.
\label{p3band}
}
\end{figure}

For the non-interacting case $U=0$, the Hamiltonian (\ref{pHub})
can be diagonalized easily and is found to have $p$ bands
$\varepsilon^\lambda(k)$. This is illustrated in Fig.~\ref{p3band}
for the case $p=3$ (the band structure for $p=2$ is shown e.g.\
in Fig.~2 of \cite{MP}). In the presence of a magnetic field
$h \ne 0$, the up and down electrons are subject to different
chemical potentials ($\mu - h/2$ and $\mu + h/2$, respectively).
A doping-dependent magnetization plateau then corresponds to
a situation where one chemical potential (say for the up electrons)
lies in one of the $p-1$ band gaps while the other (for
the down electrons) is in the middle of a band. Imposing the constraint
of fixed filling $\nc$ then requires a finite change in magnetic
field in order to move the chemical potential for the up electrons
into one of the bands next to the gap, leading to a plateau in
the magnetization curve. However, the filling of the down electrons
remains adjustable and one obtains a doping-dependent value of the
magnetization on the plateau.

A finite on-site repulsion $U > 0$ leads to corrections to
this non-interacting picture which can be treated to first order
in $U$ using standard quantum mechanical perturbation theory.
We are interested in a situation where one band for say the
up electrons is either completely filled or empty. It is therefore
sufficient to look at diagonal matrix elements of
$H_I = U \sum_{x=1}^L n_{x, \up} n_{x,\dn}$ in the basis
diagonalizing (\ref{pHub}) with $U=0$, \ie\ scattering processes
(as would be present for half-filled bands) are absent in the
present situation. This then leads to a simple shift of the critical
fields $h_{\pm}$ for a doping dependent magnetization
plateau. In particular, the plateau is non-vanishing also in the
presence of interaction at least if $U$ is small enough.

For $p=2$ this computation can be carried out explicitly (details
will be presented in \cite{prep}). One finds
that the matrix elements of $H_I$ are of a mean-field type, \ie\ simple
products of the filling of up and down electrons. The final result for the
lower and upper boundaries of the $m=1-\nc$ plateau for $p=2$ is
given by
\beq
h_{\pm} = \pm \abs{t-\tp}
    - \sqrt{t^2+\tp^2+2t \tp \cos\left(\left(2 \nc-1\right) \pi\right)}
    + (\nc-1) U + {\cal O}(U^2) \, .
\label{hcPMp2}
\eeq
This leads to a plateau width $h_{+} - h_{-}
= 2 \abs{t-\tp} + {\cal O}(U^2)$ which is independent of $U$ at first
order in $U$. The latter is a direct consequence of the mean-field form
of the matrix elements of $H_I$ and is specific to $p=2$.

We have also performed Lanczos diagonalizations of the
Hamiltonian (\ref{pHub}) for $p=2$ and $p=3$ at $U=3t$, $\tp = 0.7t$
to confirm the overall picture. Here we just mention that the
schematic Figs.~\ref{schem} are in fact based on these numerical
results, but otherwise postpone a detailed presentation to \cite{prep}.

\section{Conclusion}

To conclude, we have shown that there are two types of situations
with plateaux in the magnetization curve of a $p$-merized Hubbard
chain at arbitrary filling $\nc$: One where both $p \nc$ and $p m$
are integers and the complete excitation spectrum is gapful
and a second one where only one of the
conditions (\ref{cond}) is satisfied and part of the excitations
remain gapless. If one works at fixed filling $\nc$, the latter leads
to doping-dependent magnetization plateaux.

Although we have studied $p$-merized Hubbard chains, doping-dependent
magnetization plateaux should exist in more general situations and
indeed, similar phenomena have been theoretically observed in other systems
\cite{KLM,FrSo}.

{}From the experimental point of view, the most direct check would be
a magnetization experiment on a $p$-merized chain with controlled
doping. As we have mentioned earlier, materials realizing
dimerized and trimerized chains do exist (also at non-trivial fillings).
However, doping can be controlled better in a different class of quasi
one-dimensional materials, namely the high-$T_c$ related realizations of
$N$-leg ladders \cite{DR}. In these materials, a doping
dependence could be particularly intriguing since it could
be used to push magnetization plateaux into experimentally
accessible field regions despite the large coupling constants.
We therefore regard the magnetization process of doped $N$-leg Hubbard
ladders as an interesting field for further research.

\bigskip

{\it Acknowledgments:}
We would like to thank B.\ Dou\c{c}ot, M.\ Fabrizio, A.\ Izergin,
K.\ Le Hur, R.\ M\'elin, F.\ Mila,
T.M.\ Rice, G.L.\ Rossini and especially P.\ Degiovanni
for useful discussions and comments. D.C.C.\ acknowledges financial
support from CONICET, Fundaci\'on Antorchas, and ANPCyT (under grant
No.\ 03-00000-02249). The more involved numerical computations have
been carried out on the C4 cluster of the ETH.

\def\Jrnl#1#2#3#4{{#1}{\bf #2}, #3 (#4)}
%
%
\def\PLA{Phys.\ Lett.\ {\bf A}}
\def\PLB{Phys.\ Lett.\ {\bf B}}
\def\PR{Phys.\ Rev.}
\def\PRA{Phys\. Rev. {\bf A}}
\def\PRB{Phys.\ Rev.\ {\bf B}}
\def\PRD{Phys. Rev. {\bf D}}
\def\PRL{Phys.\ Rev.\ Lett.\ }
\def\EPJB{Eur.\ J.\ Phys.\ {\bf B}}
\def\RMP{Rev.\ Mod.\ Phys.\ }
\def\JPSJ{J.\ Phys.\ Soc.\ Jpn.\ }


\begin{thebibliography}{99}

\bibitem{AOY} M.\ Oshikawa, M.\ Yamanaka and I.\ Affleck,
{\Jrnl {\PRL}{78}{1984}{1997}}.

\bibitem{Totsuka} K.\ Totsuka, {\Jrnl {\PLA}{228}{103}{1997}};
{\Jrnl {\PRB}{57}{3454}{1998}}.

\bibitem{weSpin} D.C.\ Cabra, A.\ Honecker and P.\ Pujol,
{\Jrnl {\PRL} {79} {5126} {1997}};
{\Jrnl {\PRB}{58} {6241} {1998}}.

\bibitem{poly} D.C.\ Cabra and M.D.\ Grynberg,
{\Jrnl {\PRB}{59} {119} {1999}}; A.\ Honecker,
{\Jrnl {\PRB}{59} {6790} {1999}}.

\bibitem{trimer} M.\ Ishii {\it et.al.},
{\Jrnl {\JPSJ}{69} {340} {2000}}.

\bibitem{NH4CuCl3} W.\ Shiramura {\it et.al.},
{\Jrnl {\JPSJ}{67} {1548} {1998}}.

\bibitem{KLM} H.\ Tsunetsugu, M.\ Sigrist and K.\ Ueda.
{\Jrnl{\RMP}{69}{809}{1997}}.

\bibitem{FrSo} H.\ Frahm and C.\ Sobiella,
{\Jrnl{\PRL}{83}{5579}{1999}}.

\bibitem{DR} E.\ Dagotto and T.M.\ Rice,
{\Jrnl {Science }{271} {618} {1996}}.

\bibitem{orgS} J.T.\ Devreese, R.P.\ Evrard and V.E.\ van Doren (eds.),
{\it Highly Conducting One-Dimensional Solids}, Plenum Press, New York (1979);
T.\ Ishiguro and K.\ Yamaji, {\it Organic Superconductors},
Springer Series in Solid-State Sciences 88, Berlin (1990).

\bibitem{EIT} T.\ Egami, S.\ Ishihara and M.\ Tachiki,
{\Jrnl {Science }{261} {1307} {1993}}.

\bibitem{LW} E.\ Lieb and F.Y.\ Wu,
{\Jrnl {\PRL} {20} {1445} {1968}}.

\bibitem{FK} H.\ Frahm and V.E.\ Korepin,
{\Jrnl {\PRB}{42} {10553} {1990}};
{\Jrnl {\PRB}{43} {5653} {1991}}.

\bibitem{EF} F.H.L.\ E{\ss}ler and H.\ Frahm,
{\Jrnl {\PRB}{60} {8540} {1999}}.

\bibitem{PS} K.\ Penc and J.\ Solyom, {\Jrnl {\PRB}{47} {6273} {1993}}.

\bibitem{gen2} A.O.\ Gogolin, A.A.\ Nersesyan and A.M.\ Tsvelik,
{\it Bo\-so\-ni\-za\-tion and Strongly Correlated Electron Systems},
Cambridge University Press, Cambridge (1998).

\bibitem{MP} K.\ Penc and F.\ Mila, {\Jrnl {\PRB} {50} {11429} {1994}}.

\bibitem{Tan} K. Tandon {\it et.al.},
{\Jrnl {\PRB}{59} {396} {1999}}.

\bibitem{Kagome}
 P.\ Lecheminant {\it et.al.}, {\Jrnl {\PRB}{56} {2521} {1997}};
 Ch.\ Waldtmann {\it et.al.}, {\Jrnl {\EPJB}{2} {501} {1998}}.

\bibitem{FGN} M.\ Fabrizio, A.O.\ Gogolin and A.A.\ Nersesyan,
{\Jrnl {\PRL}{83} {2014} {1999}}.

\bibitem{prep} D.C.\ Cabra, A.\ De Martino, A.\ Honecker,
  P.\ Pujol and P. Simon, in preparation.

\end{thebibliography}
\end{document}